\documentstyle[12pt,epsf,epsfig,wrapfig]{article}  
\begin{document}  
\newtheorem{thm}{Theorem}
\newtheorem{cor}{Corollary}
\newtheorem{Def}{Definition}
\begin{center}  
{\large \bf  
Bell's inequality, the Pauli exclusion principle and baryonic
structure
\\ }  
\vspace{5mm}  
Paul O'Hara  
\\  
\vspace{5mm}  
{\small\it  
Dept. of Mathematics, Northeastern Illinois University, 5500  
North St. Louis Avenue, Chicago, IL 60625-4699, USA\\}  
\end{center}  
%
%
%
\begin{abstract}
Bell's inequality has been traditionally used to explore the
relationship between hidden variables and the Copenhagen
interpretation of quantum mechanics.  In this paper, another use
is
found. Bell's inequality is used to derive a coupling principle
for
elementary particles and to give a deeper understanding of
baryonic
structure.  We also give a derivation of the Pauli exclusion
principle from the coupling principle.
Pacs: 14.20-c, 12.90+b, 3.65.Bz, 12.40.Ee   
\end{abstract}

\section {Introduction}

Bell's Inequality \cite{bell} was derived by John Bell in 1964 as
a response
to The Einstein-Podolsky-Rosen Paradox \cite{ein}, a problem
pertaining to the foundations of quantum physics. 
Bell saw his inequality as being able to discern between two
different epistemological views of quantum mechanics, the one
proposed by EPR and the one proposed by the Copenhagen
interpretation of quantum theory.   

In this paper we point out another implication of Bell's
work. We first derive a coupling principle directly from the
inequality and show that the Pauli principle can be viewed as a
special
case of this coupling. We then apply the principle to further our
understanding of baryonic structure and note that the case of
spin $\frac32$ baryons can be analyzed in one of two ways as
reflected in the following assumptions:\newline
(1) In every direction the spin will be observed to be $\pm 3/2$.
\newline
(2)There exists some direction in which the spin will be observed
to be $\pm 3/2$.

Assumption (1) in fact is the key point of a previous paper
\cite{spin} and will not be discussed here. Assumption (2), on
the other hand, when combined with the coupling  principle
mentioned above, enables us to explain the statistical structure
of the $\Delta^{++}$ and the $\Omega^-$ particles without any
recourse to color.  It is discussed in section four of the paper.

\section{A Coupling Principle}

Consider three (or more) particles in  
the same spin state. In other words, if a measurement is  
made in an arbitrary direction $a_1$ on ONE of the three  
particles, then the measurements can be  
predicted with certainty for the same direction for each of the
other particles. We point out immediately that such spin
correlations are isotropic for the particles under discussion and
that we are not dealing with a 
polarization phenomena where spin correlations exist for a  
preferred direction.  In our case, the particles are spin-  
correlated in all directions at once, as for example in the case 
of two particles in a singlet state.  Hence, the
initial
direction of measurement is arbitrary. We refer to such particles
as isotropically spin-correlated particles.\cite{epr}   
  
Specifically, if we denote a spin up state by the ket $|+>$ and a
spin down state by the ket $|->$ then without loss of generality,
we can assume that the three particles have the joint spin state
$|+, +, +>_1$ ($|-, -, ->_1$), where the suffix 1, refers to the
observed spin states in 
the arbitrary direction $a_1$.   

In the  
language of probability, we can say that if the spin state of a  
particle is $|+>_1$ then the corresponding spin state of each of 
the other two particles can be predicted (for the same direction)
with probability 1.  Furthermore, the probability 1 condition  
means that in principle spin can now be measured simultaneously  
in the three  
different directions $a_1, a_2, a_3$, for the three particle  
ensemble (see Fig.  
1). Let $P$ denote the joint probability measure relating the  
measurements in the three different directions and recall the  
fact that if spin is  
observed to be in the $|+>_1$ state in direction $a_1$ for one of
the particles then the  
conditional probability of  
observing $|+>_2$ or $|->_2$ in the direction $a_2$ for a second 
particle, is given by  
$\cos^2(c\theta_{12})$ or $\sin^2(c\theta_{12})$ respectively, 
where $\theta_{12}$ is the angle subtended by $a_1$ and $a_2$ and
$c$ is a constant. For the purpose of the argument below, we will
work with $c=1/2$. However, the argument can be made to work for
any
value of $c$, and in a particular way can be applied to the spin
of a photon, provided $c=1$. 

\begin{wrapfigure}{r}{8cm}  
\epsfig{figure=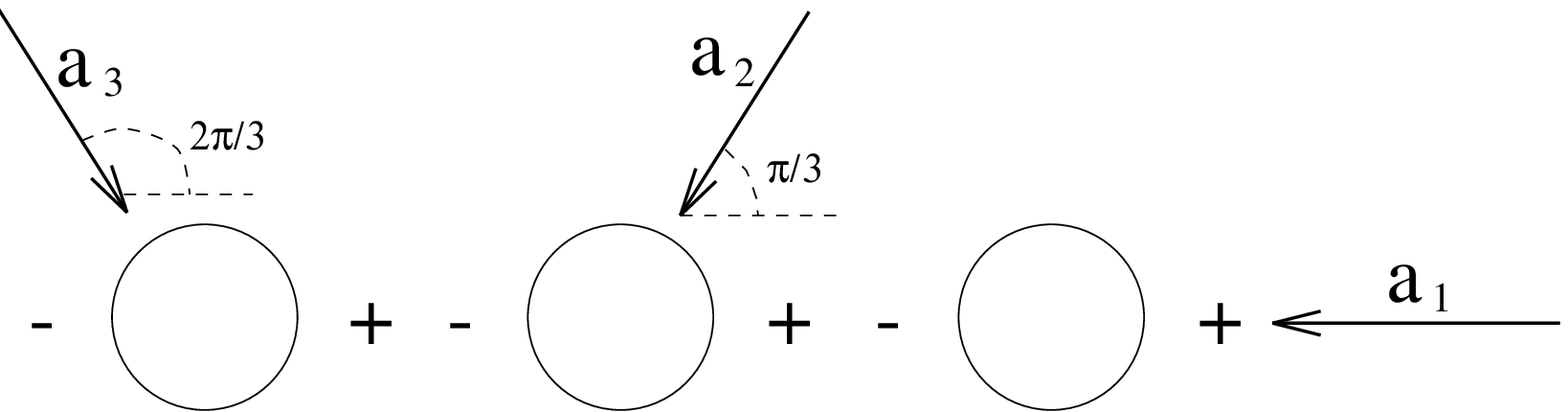,width=8cm}  
{\small Figure 1: Three isotropically spin-correlated
particles.}
\end{wrapfigure}  
  
With notation now in place, we adapt an argument of Wigner
\cite{wig} to
show that isotropically spin-correlated particles must occur in
pairs.  We prove this by contradiction. Specifically, consider
three isotropically
spin-correlated
particles (see Fig. 1), as explained above.  It follows from the 
probability 1 condition,that three  
spin measurements can be  
performed, in principle, on the three particle system, in the  
directions  $a_1,\ a_2, \ a_3$. Let  
$(s_1, s_2, s_3)$ represent the observed spin  
values  
in the three different directions. Note  
that $s_i = \pm $ in the notation developed above   
which means that there exists only two possible values for each  
measurement. Hence,for three measurements there are a total of 8 
possibilities in total.  
In particular,  
  
\vspace{-6mm}  
  
\begin{eqnarray*}  
\;\;\;\;\;\;\{(+,+,-), (+,-,-)\} & \subset &  
\{(+,+,-),(+,-,-),(-,+,-),(+,-  
,+)\}\\  
\Rightarrow \;\;\; P\{(+,+,-),(+,-,-)\} & \le &  
P\{(+,+,-),(+,-,-),(-  
,+,-),(+,-,+)\}.  
\end{eqnarray*}  
Therefore,  
$$\frac 12 \sin^2 \frac {\theta_{31}}{2}\le \frac 12 \sin^2 \frac
{\theta_{23}}{2} + \frac 12 \sin^2 \frac {\theta_{12}}{2}.$$  If
we
take  
$\theta_{12}=\theta_{23}=\frac {\pi}{3}$ and  
$\theta_{31}=\frac {2\pi}{3}$ then this gives $\frac 12 \ge \frac
34$ which is clearly a contradiction. In other words, three  
particles cannot all be in the same spin state with probability  
1, or to put it another way, isotropically spin-correlated
particles must occur in pairs. 

Finally, as noted above, this argument applies also to spin 1
particles, like the photons, provided full angle formulae are
used instead of the half-angled formulae. 

\section{Pauli exclusion principle}

The above results can be cast into the form of a theorem (already
proven above) which
will be referred to as the ``coupling principle''.
\begin{thm}(The Coupling Principle) Isotropically
spin-
correlated particles must occur in PAIRS.\end{thm} 

In practice, isotropically spin-correlated particles occur when 
the particles' spin are either anti-parallel (singlet
state) or parallel to each other. 

We now show that when a system of indistinguishable particles
contain ``coupled'' particles then this system of particles must
obey fermi-dirac statistics.  We first do this for a 2-
particle spin-singlet state system and then extend the result to
an n-particle
system.  Throughout  $\lambda_i=(q_i, s_i)$ will represent the
quantum coordinates of particle i, with $s_i$ referring to the
spin coordinate and $q_i$ representing all other coordinates. In
practice, 
$\lambda_i=(q_i, s_i)$ will represent the eigenvalues of an
operator defined on the Hilbert space $L^2({\cal R}^3)\otimes
H_2$,
where 
$H_2$ represents a two-dimensional spin space  of particle i.  
We will mainly work with $\lambda_i$. However, occasionally, in
the interest of clarity,
we will have need to distinguish the $q_i$ from the $s_i$.  
\begin{cor} Let $|\psi (\lambda_1,\lambda_2)>$ denote
a two particle state where the $\lambda_1$ and
$\lambda_2$ are as defined above.
If the particles are in
a spin-singlet state then their joint state function  will be
given by
$$|\psi_i(\lambda_1,\lambda_2)>=\frac{1}{\sqrt
2}[|\psi_1(\lambda_1
)>  \otimes
|\psi_2(\lambda_2)>-
|\psi_1(\lambda_2)> \otimes |\psi_2(\lambda_1)>].$$  
In other words, coupled particles obey fermi-dirac
statistics.\end{cor}  
{\bf Proof:} The general form of the two particle eigenstate is
of the form
$$|\psi(\lambda_1,\lambda_2)>=c_1|\psi_1(\lambda_1)> \otimes
|\psi_2(\lambda_2)>+c_2|\psi_1(\lambda_2)> \otimes
|\psi_2(\lambda_1)>.$$
Since the particles are in a spin-singlet state then $P(\lambda_1
=\lambda_2)\le P(s_1=s_2)=0$. Therefore, $<\psi(\lambda_1,
\lambda_1)|\psi(\lambda_1, \lambda_1)>=0$ and hence
$|\psi(\lambda_1, \lambda_1)>=0$, from the inner product
properties of a Hilbert space.
It follows, that $c_1=-c_2$ when the particles are
coupled and normalizing the wave function gives $|c_1|=\frac{1}{\sqrt{2}}$.
The result follows. 
QED 

Note that the same result can also be used to describe particles
whose spin correlations are parallel to each other in each
direction. This can be done by correlating a measurement in
direction $a$ on one particle, with a measurement in direction $-
a$ in the other.  In this case, the state vector for the parallel
and anti-parallel measurements will be found to be by the above
argument: 
$$|\psi(\lambda_1,\lambda_2)>=\frac{1}{\sqrt
2}[|\psi_1(\lambda_1)>
\otimes
|\psi_2(\lambda_2(\pi))>-|\psi_1(\lambda_2)> \otimes
|\psi_2(\lambda_1(\pi))>]$$
where the $\pi$ expression in the above arguments, refer to the
fact that the measurement on particle two is made in the opposite
sense, to that of particle one.

This result can now be generalized to derive the Pauli exclusion
principle for a system of n indistinguishable particles
containing an least one pair of coupled particles.
First, note the following use of notation. 
Let $|\psi_i(\lambda_j)>=\psi_i(\lambda_j)\vec e$ where
$\psi_i(\lambda_j)$ refers to particle $i$ in the state 
$|\psi_i(\lambda_j)>$ and $\vec e$ is a unit vector. Then 
\begin{eqnarray*}
|\psi_i(\lambda_j)>\otimes 
|\psi_k(\lambda_l)>&=&[\psi_i(\lambda_j)\vec e_1]\otimes[\psi_k
(\lambda_l)\vec e_2]=\psi_i(\lambda_j)\psi_k(\lambda_l)\vec e_1
\otimes \vec e_2\\&=&|\psi_k(\lambda_l)>\otimes
|\psi_i(\lambda_j)>.\end{eqnarray*}
In other words, the tensor product is commutative.
From now on we will drop the ket notation
and simply write that $\psi_i(\lambda_j)\otimes \psi_k(\lambda_l)
= \psi_k(\lambda_l) \otimes \psi_i(\lambda_j)$, with ket
notation being understood. We also denote an n-particle state
by $\psi_{1\dots n}[\lambda_1, \dots , \lambda_n]$ where the
subscript $1 \dots n$ refer to the n particles. However, when
there is no ambiguity involved we will simply write this n-
particle state as $\psi[\lambda_1, \dots , \lambda_n]$ with the 
subscript $1\dots n$ being understood.  Finally,  note that for
an indistinguishable system of n particles
$$\psi [\lambda_1,\dots ,\lambda_n]
=\sum_P \sigma_P c_P\psi(\lambda_1, \dots ,\lambda_n)$$ where  
$\psi(\lambda_1, \dots ,\lambda_n)=
\psi_{1}(\lambda_1)\otimes
\dots \otimes \psi_{n}(\lambda_n)$  and 
$\sigma_P(\psi_1\otimes \dots \otimes \psi_n)=\psi_{i_1} \otimes
\dots \otimes \psi_{i_n}$, gives a permutation of the states.
With this notation, we now prove the following theorem:
\begin{thm} (The Pauli Exclusion Principle) A
sufficient condition for a system of n indistinguishable
particles
to exhibit fermi-dirac statistics is that it contain spin-coupled
particles .\end{thm}
{\bf Proof:}  We will work with three particles, leaving the
general case for the Appendix.  Consider a system of three
indistinguishable particles, containing spin-coupled particles.
Using the above notation and applying
Cor 1 in the second line below, we can write:
\begin{eqnarray*} \psi [\lambda_1, \lambda_2,
\lambda_3]&=&\frac{1}{\sqrt 3}
\{\psi_1(\lambda_3)\otimes\psi_{23}[\lambda_1,\lambda_2] +
\psi_2(\lambda_3)\otimes\psi_{31}[\lambda_1,\lambda_2]+
\psi_3(\lambda_3)\otimes\psi_{12}[\lambda_1,\lambda_2]\}\\
&=&\frac{1}{\sqrt {3!}}\{\psi_1(\lambda_3)\otimes [\psi_2(\lambda_1)\otimes
\psi_3(\lambda_2)-
\psi_3(\lambda_1)\otimes \psi_2(\lambda_2)]\\ 
\ & &+\psi_2(\lambda_3)\otimes [\psi_3(\lambda_1)\otimes
\psi_1(\lambda_2)-
\psi_1(\lambda_1)\otimes \psi_3(\lambda_2)]\\ 
\ &&+\psi_3(\lambda_3)\otimes [\psi_1(\lambda_1)\otimes
\psi_2(\lambda_2)-
\psi_2(\lambda_1)\otimes \psi_1(\lambda_2)]\}\\
&=&\sqrt{3!}\psi_1(\lambda_1)\wedge \psi_2(\lambda_2)\wedge
\psi_3(\lambda_3)
\end{eqnarray*} where $\wedge $ represents the wedge product.
Thus the
wave function for the three indistinguishable particles obeys the
fermi-dirac statistics.  The n-particle case follows by
induction. QED.

Mathematically it is possible to give other reasons why
$P(\lambda_i, \lambda_i)=0$ (quark ``color'' being a case in
point)
In fact, a necessary and sufficient condition
can be formulated for
fermi-dirac statistics as follows: In a system of n-
indistinguishable particles $\psi [\lambda_1, \dots \lambda_i,
\lambda_i,\dots]=0$ for
the $i$ and $j$ states if and only if 
$$\psi [\lambda_1, \lambda_2,
\dots \lambda_n]=\sqrt{n!}\psi_1(\lambda_1)\wedge
\psi_2(\lambda_2)\wedge
\dots \wedge
\psi_n(\lambda_n).$$
The sufficient part of the proof will be the
same as in Theorem 2 while the necessity part is immediate. 
However, the significance of Theorem 2 lies in the fact that
for spin-type systems, particles may couple and  this
coupling causes fermi-dirac statistics to occur.
Moreover, the coupling would appear to be a more universal
explanation of the Pauli exclusion principle than for example
``color''. Not only does it explain the statistical structure of
the baryons (see below) but it also explains why in chemistry
only two electrons
share the same orbital and why ``pairing''  occurs in the theory
of superconductivity.\cite{epr},\cite[p8]{ohara}
 
\vspace{3mm}    

\section {Baryonic Structure}

We now apply the coupling principle to shed further understanding
on the standard model interpretation of baryonic structure. 
However, first we need to agree on the meaning of
spin 1/2 and spin 3/2 particles. 

We say that a particle has spin 1/2 if an ensemble 
of such particles decomposes into two spin states when passed
through a Stern-Gerlach magnet. The respective spin states will
be referred to 
as spin 1/2 and spin -1/2 and each occur with probability 1/2. 
We say that a particle has spin 3/2 if an ensemble of such
particles decomposes into four spin states when passed through a
Stern-Gerlach  magnet, states which correspond to spin 3/2, 1/2,
-1/2, -3/2. Moreover, this in effect means that there exists some
direction in which the spin will be $\pm 3/2$ and is consistent
with assumption (2) of the introduction. 

Next, we apply the coupling principle to spin 1/2 baryons.  We
note that 
if two quarks are coupled in a singlet state then the
remaining quark 
is statistically independent of the other two and can exhibit
spin 1/2 or 
spin -1/2 properties. It follows that the spin 1/2 baryon can be
explained 
by the coupling principle if two of the three particles are in a
singlet 
state. Moreover, if in addition the three quarks are
indistinguishable then it follows from the Pauli
principle (Theorem 2) that the wave function can be
expressed as:
$$\psi[\lambda_1,\lambda_2,\lambda_3]=\frac{1}{\sqrt3}[\psi_{12}[
\lambda_1,\lambda_2]\psi_3(\lambda_3)+\psi_{31}[\lambda_1,\lambda
_2]\psi_2(\lambda_3)
+\psi_{23}[\lambda_1,\lambda_2]\psi_1(\lambda_3)]$$ 
where
$\psi_{ij}[\lambda_1,\lambda_2]=\frac{1}{\sqrt2}(\psi_i(\lambda_1
)\psi_j(\lambda_2)-
\psi_i(\lambda_1)\psi_j(\lambda_2))$.
In other words, fermi-dirac statistics results.

The case of spin 3/2 baryons can be explained in a similar way.  
In this case, it is sufficient to choose the coupled particles in
the parallel-correlated-spin state (spin +1 or spin -1). The
third particle will 
have a spin statistically independent of the coupled particles. 
Putting the three particles together now gives the baryon of spin
3/2, in the sense above. Its state can be formally written as:
$$c(\psi(\lambda_1,
\lambda_2)_{12}\psi_3(\lambda_3)+\psi(\lambda_1,
\lambda_2)_{31}\psi_2(\lambda_3)+
\psi_{23}(\lambda_1, \lambda_2)\psi_1(\lambda_3)$$ 
where
$\psi_{ij}(\lambda_1,\lambda_2)=\psi_i(\lambda_1)\psi_j(\lambda_
2)+
\psi_i(\lambda_2)\psi_j(\lambda_1)$.
and $c=1/3$ if $q_1=q_2=q_3$, $c=0$ if $\lambda_1=\lambda_2$, 
$c=\frac{1}{\sqrt3!}$ otherwise.

A couple of things to note:\newline
(1) The fact that two of the quarks are coupled and the other is
independent might suggest that the particles can be identified.
However, indistinguisability in effect prohibits this. An
analogous situation occurs in chemistry when molecules in a
liquid form a dynamic equilibrium, and give rise to a constant
association and disassociation of the ions, in accordance with
indistinguisability. This suggests that the quarks
are in effect in dynamic equilibrium with each other, with the
coupling being continuously broken and then reformed among the
different quarks.\newline
(2) If it is assumed that only singlet state coupling 
(in the sense of Theorem 1) 
is stable then all spin 3/2 baryons will necessarily decompose.
This conjecture would seem to be substantiated by the very rapid
decay rates observed in ALL spin-3/2 baryons.\newline 
(3) It is impossible for all three particles to be isotropically 
spin-correlated.\newline
(4) By using the coupling principle, the principle of isotropy
still applies. The particle does not have a preferred orientation
in space.\newline 
(5) In the case of three statistically independent particles, the
joint state will be similar in form to the expression for
parallel coupling. However, the constant coefficients will be
different and the state will have the form:
$$c(\psi(\lambda_1,
\lambda_2)_{12}\psi_3(\lambda_3)+\psi(\lambda_1,
\lambda_2)_{31}\psi_2(\lambda_3)+
\psi_{23}(\lambda_1, \lambda_2)\psi_1(\lambda_3)$$ 
where
$\psi_{ij}(\lambda_1,\lambda_2)=\psi_i(\lambda_1)\psi_j(\lambda_2
)+\psi_i(\lambda_2)
\psi_j(\lambda_1)$,
and $c=\frac{1}{\sqrt 8}$ if $q_1=q_2=q_3$, and
$c=\sqrt\frac{3}{8}$ otherwise. 
  
\vspace{3mm}

\section {Conclusion:} 
In general, we can conclude that  
isotropically spin-correlated  
particles, can only occur in pairs. This  
suggests a ``spin-coupling principle''.  It follows that three or
more such correlated particles cannot exist and in particular,  
the  
structure of spin-$\frac32$ baryons cannot be composed of three  
isotropically correlated quarks, without giving rise to a  
mathematical  
contradiction. However, their statistical structure may be
explained in terms
of the coupling principle, without any recourse to the concept of
color.  Finally, we note that the Pauli exclusion  
principle can be derived from the ``spin coupling principle''.   

\section {Appendix -- Alternate Proof of Pauli Exclusion
Principle}   

{\bf Theorem 2:} {\it A sufficient condition for a system of n
indistinguishable particles to exhibit fermi-dirac statistics is
that it contain spin-coupled particles.}

\noindent {\bf Proof:} Let $\psi [\lambda_i, \lambda_j]$
represent the
coupled state. From the coupling principle we get
$$0\le P(\lambda_1, \dots , \lambda_i \dots \lambda i \dots
\lambda_n)\le P(\lambda_i=\lambda_j) \le P(s_i=s_j)=0$$
Therefore 
\begin{equation}
\psi [\lambda_1, \dots \lambda_i \dots \lambda_i \dots
\lambda_n]=0
\label{a1}
\end{equation}
The general form of the wave function is given by
\begin{eqnarray*} \psi [\lambda_1, \dots \lambda_n]&= \sum_P
\sigma_P c_p
\psi_1(\lambda_i) \otimes \psi_2(\lambda_j) \otimes \dots \otimes
\psi (\lambda_n)\\
&= \sum_P \sigma_P d_p \psi_1(\lambda_j) \otimes
\psi_2(\lambda_i)\dots \otimes \psi (\lambda_n)
\end{eqnarray*}
where $c_P$ and $d_P$ are constants.  
Indistinguisability implies that $c_P^2=d_P^2=\frac {1}{n!}$, for
each $P$. Let $c=\frac {1}{\sqrt n!}$.
If we now put $\lambda_i=\lambda_j$ then we get from equation
(\ref{a1})  that
$$0=\sum_P(c_P+d_P)\sigma_P
\psi_1(\lambda_i)\otimes\psi_2(\lambda_i)\otimes \dots \otimes
\psi_n(\lambda_n)$$
Therefore, $c_P=-d_P$ for each $P$ by the linear independence of
the eigenfunctions. It follows, 
\begin{eqnarray*} &\ &\psi[\lambda_1, \dots
,\lambda_n]\\
&=&(c/2)\sum_P\sigma_P\bigl (\psi_1(\lambda_i)\otimes
\psi_2(\lambda_j)\otimes \dots \otimes \psi_n(\lambda_n)-
\psi_1(\lambda_j)\otimes\psi_2(\lambda_i)\otimes \dots \otimes
\psi_n(\lambda_n)\bigr )\\
&=&\sqrt{n!}\psi_1(\lambda_1)\wedge \dots \psi_n(\lambda_n).
\end{eqnarray*}
This obeys the fermi-dirac statistics and the result follows. QED

\vspace{3mm}

\noindent{\bf Acknowledgments:} I would like to thank the
``Foundations of
Quantum Mechanics Group" of the University of Chicago, Nick
Huggett of
University of Illinois at Chicago, Oleg Taryaev of Dubna, and the PRL 
referee for their
very useful
comments and suggestions.

\end{document}